\documentclass[12pt,preprint]{aastex}

\newcommand{\kms}{\mbox{km\,s$^{-1}$}}

\newcommand{\fdeg}{\mbox{\rlap{.}$^\circ$}}

\slugcomment{submitted to ApJ}

\shorttitle{Bipolar outflow in IRAS 08544$-$4431}
\shortauthors{Dinh-V-Trung}

\begin{document}

\title{The peculiar molecular envelope around the post-AGB star IRAS 08544$-$4431}

\author{Dinh-V-Trung\footnote{on leave from Center for Quantum Electronics, Institute of Physics,
Vietnamese Academy of Science and Technology, 
10 DaoTan, BaDinh, Hanoi, Vietnam}}
\affil{Institute of Astronomy and Astrophysics, Academia Sinica, P.O. Box 23-141, Taipei 106, Taiwan}
\email{trung@asiaa.sinica.edu.tw}

\begin{abstract}
Circumbinary disks have been hypothesized to exist around a number of binary post-AGB stars. Although
most of the circumbinary disks have been inferred through the near IR excess, a few of them are strong
emitters of molecular emission. Here
we present high angular resolution observations of the emission of $^{12}$CO and 
its isotopomer $^{13}$CO J=2--1 line from the circumstellar envelope around the binary post-AGB star 
IRAS 08544$-$4431, which is one of the most prominent members of this class of objects. 
We find that the envelope is resolved in our observations and two separate components can 
be identified: (a) a central extended and strong component with very narrow linewidth between 2 - 6 \kms;
(b) a weak bipolar outflow with expansion velocity up to 8 \kms.
The central compact component possesses low and variable $^{12}$CO/$^{13}$CO J=2--1 line ratio, indicating
optically thick emission of the main isotope. We estimate a molecular gas mass of 0.0047 M$_\odot$ for
this component based on the optically thinner $^{13}$CO J=2--1 line.
We discuss the relation of the molecular envelope and the circumbinary disk inferred from near IR excess
and compare with other known cases where the distribution of molecular gas has been imaged at high
angular resolution. 

\end{abstract}


\section{Introduction}

The rapid evolution of low and intermediate mass stars (0.5 to 8 M$_\odot$) after the end of the Asymptotic 
Giant Branch (AGB) phase is accompanied by a radical change in morphology of the circumstellar envelope
around them. The circumstellar envelope created by the slow and dusty stellar wind during the AGB phase is
known to be roughly spherically symmetric as the radiation pressure on dust grains is expected
to be isotropic. However, a variety of morphologies ranging from spherical to multipolar shapes have been
observed in post-AGB envelopes and planetary nebulae (Sahai et al. 2007). 
This morphological change has also been seen together
with the appearance of collimated high velocity outflows such as the Egg nebula (Cox et al. 2000) or CRL 618
(S\'{a}nchez-Contreras et al. 2004). The origin of the morphology and the mechanism to generate the high velocity
outflow are still under active study. One commonly suggested mechanism is the presence of a binary 
companion (Balick \& Frank 2002).
The companion can capture wind material into a rotating disk around it. Consequently, high velocity jets 
could be launched from the disk through magneto-centrifugal effect (Soker 2006, Nordhaus \& Blackman 2006,
Nordhaus et al. 2007). Alternatively, the gravitational force of the companion 
may also attract the wind material toward the equatorial, thus forming a circumbinary disk-like structure
or torus. Numerical simulations by Mastrodemos \& Morris (1999) and more recently by Edgar et al. (2008)
suggest that the torus is likely to be in expansion and not in rotation around the central star. However,
when the torus is dense enough, it could serve to confine and channel the wind from the 
central star into bipolar directions.

The presence of rotating disks around a large number of post-AGB stars has been inferred from the IR excess 
(De Ruyter et al. 2006). The excess suggests that hot dust exists close to the hot post-AGB stars, even 
though the mass loss due to dusty slow wind is expected to have ceased long ago. However, 
until recently only the rotating disk around post-AGB star Red Rectangle has been imaged at high 
angular resolution (Bujarrabal et al. 2005).

IRAS 08544$-$4431 is a F-type post-AGB star and is among the first post-AGB 
stars discovered to have near IR excess (Maas et al. 2003). From radial velocity measurements, Maas et al. (2003)
inferred that the star is a binary system with an orbital period of
499 days and a mass function, which is a measure of the ratio between the companion mass to the primary mass
in a single-line spectroscopic binary system, of 0.02 M$_\odot$. 
Using a typical luminosity for post-AGB star, de Ruyter et al. (2006)
suggested a distance close to 1 kpc for IRAS 08544$-$4431. Spectroscopic observations by Maas et al. (2003) 
show that the star is an O-rich post-AGB star. More interesting, the abundance of metals 
with high condensation temperature is strongly depleted. Such anormaly could be
explained as gas and dust separation and reaccretion from a long-lived reservoir of material around the star, i.e
a rotating disk. CO emission lines have also been detected from IRAS 08544$-$4431 (Maas et al. 2003). 
Unlike the parabolic shape seen in the envelope around AGB stars, the line
profiles of CO lines of IRAS 08544$-$4431 are very peculiar with a strong central peak and extended wings, 
suggesting unusual kinematics of the molecular gas in the envelope.
Using optical interferometry technique, Deroo et al. (2007) resolved the near IR continuum emission into an
elongated structure. They found that both the SED and the interferometric data 
could be well fitted using the irradiated dust disk model. However, information on the kinematics of the molecular gas 
around IRAS 08544$-$4431 is really needed to settle the question concerning the existence of a rotating disk 
around this star.

In this paper we present high angular resolution observations of CO J=2--1 transition and its isotopes
from the envelope around IRAS 08544-4431 using the Sub-millimeter array. We resolve the envelope and probe
its morphology and kinematics. In Sec. 4 we will discuss the implication of our new results.

\section{Observation}
We use the Sub-Millimeter Array (SMA), which consists of 8 antennas of
6 m in diameter, to observe IRAS 08544$-$4431. The observation was carried out
during the night of 09 April 2007 under excellent weather conditions.
The zenith opacity of the atmosphere at 230 GHz was around
0.1, resulting in antenna temperatures (single sideband) in the range
of 300 K to 400 K. In our observation the SMA provides projected baselines in the range
between 6 m to 78 m. The total on-source integration time of our 
observation is slightly less than 03 hours. 
The coordinates of IRAS 08544$-$4431 taken from Maas et
al.\ (2005), $\alpha_{\rm J2000}$=08:56:14.18, $\delta_{\rm
J2000}$=$-$44:43:10.7, were used as phase center in our
observation. The nearby and relatively strong quasar 0826$-$225 
was monitored frequently to correct
for gain variation due to atmospheric fluctuations. Saturn and its moon
Titan were used as bandpass and flux calibrator,
respectively. The large bandwidth ($\sim$2 GHz) of the SMA correlator
allows us to cover simultaneously the $^{12}$CO $J$=2--1 line in the
upper sideband and the $^{13}$CO $J$=2--1 and C$^{18}$O $J$=2--1
lines in the lower sideband. In our observation the SMA correlator was
setup in normal mode, providing a frequency resolution of 0.825 MHz or
$\sim$1 \kms\ in velocity resolution.  The visibilities are edited and
calibrated using the MIR/IDL package (Scoville et al. 1993), which is developed specifically
for SMA data reduction. The calibrated data are then exported for
further processing with the MIRIAD package (Sault et al. 1995).  The line data are then
Fourier-transformed to form dirty images. Deconvolution of the dirty
images is done using the task $clean$.  The resulting synthesized beam
for $^{12}$CO $J$=2--1 channel maps is 5\farcs9x2\farcs2 at position
angle PA=17\fdeg5. The corresponding conversion factor between flux
and brightness temperature is 1.76 K Jy$^{-1}$.  The rms noise level for each
channel of 1 \kms\ is 110 mJy beam$^{-1}$. $^{13}$CO $J$=2--1
emission has been also detected and imaged. The synthesized beam for the
$^{13}$CO J=2--1 emission is 5\farcs8x2\farcs4.  
The C$^{18}$O J=2$-$1 line is not detected in our observations and, therefore, 
we will not discuss this line any further in this paper.

We also form the continuum image by averaging line free channels in the
upper sideband. No 230 GHz continuum emission from IRAS 08544$-$4431 is detected in our
observations with rms noise level of 3 mJy beam$^{-1}$.

\section{Results}
In Figure 1, we show the channel maps of $^{12}$CO J=2--1 emission. Because of the very low
declination of the source, resulting in an elongated synthesized beam, the emission in 
the channel maps near the systemic velocity appears extended and resolved. 
The deconvolved size of the emission at velocity
V$_{\rm LSR}$ = 44 km s$^{-1}$ is 4".3x4".5. The $^{12}$CO J=2--1 emission also reveals
more asymmetry to the North East of the envelope, namely in channels at velocity 
V$_{\rm LRS}$ = 45 - 46 \kms. 
In Figure 3 we compare the total flux of the
$^{12}$CO J=2--1 line detected with the SMA with that obtained previouly with SEST
telescope (Maas et al. 2003). The line shape is very similar between interferometric and
single dish observations, although the line intensity seen by the SMA is slightly higher,
probably related to the difficulty of calibrating the absolute flux for this very low
declination source. Therefore we conclude that
the SMA detects all the flux of the $^{12}$CO J=2--1 line.

In Figure 2 we show the channel maps of $^{13}$CO J=2--1. This line is significantly 
weaker than the J=2--1 line of the main isotope and its emission appears spatially less
extended. We note that
the $^{13}$CO J=2--1 line is also noticeably narrower, covering a velocity range between
42 - 48 \kms, as can be seen in the total intensity profiles shown in Figure 3.
Only central velocity channels between 44 and 46 \kms\, contain significant amount of emission. 
The higher velocity component seen in the main isotope line is not detected in $^{13}$CO J=2--1 emission.
That suggests a strong variation in the $^{12}$CO/$^{13}$CO line ratio within the envelope of IRAS
08544$-$4431. We will discuss further about the implication of the variation of this ratio in the following
section.  

Toward the wings the $^{12}$CO J=2--1 emission becomes more compact. In addition, the emission 
at the redshifted and blueshifted
wings show slight positional velocity shift. In the redshifted wing, the emission is located
to the North West of the center. Whereas, the centroid of the emission in the blueshifted wing is
slightly shifted toward the South East (see Fig. 1). Such trend in spatial kinematics of the 
emission could be more easily seen in the position-velocity diagram of the cuts along the position
angle PA=19$^\circ$ and the perpendicular direction at PA=109$^\circ$, which are shown in 
Figure 4. A slight position-velocity gradient can bee seen in the digram along the minor axis of the envelope. 
That position-velocity gradient, which is often seen in post-AGB envelopes and
young planetary nebulae with high velocity jets such as CRL 2688 (Cox et al. 2000),
CRL 618 (Sanchez Contreras et al. 2004) and NGC 6302 (Dinh-V-Trung et al. 2008), indicates 
the presence of a weak bipolar outflow in the
envelope of IRAS 08544$-$4431. 

From the properties of the envelope discussed above, we tentatively separate the envelope 
into two kinematic components: an extended and massive envelope, having a very narrow linewidth of
about 2 to 6 \kms, and a weak bipolar outflow with expansion velocity up to 8 \kms.

\section{Molecular gas in IRAS 08544$-$4431}
\subsection{CO line opacity}

In IRAS 08544$-$4431 the $^{12}$CO $J$=2--1 transition is very likely optically
thick. This is supported by two observational results: the relatively
intense $J$=1--0 line and the relatively high and variable
$^{12}$CO/$^{13}$CO $J$=2--1 intensity ratio. 

The $^{12}$CO $J$=1--0 transition was observed by Maas et al. (2003) using the
same SEST telescope. The peak main beam temperature is $\sim$0.15 K. 
Because the emitting region as seen in our SMA channel maps
is small in comparison to the telescope beams, when converted to the
same telescope beam, the 
intensity ratio of $^{12}$CO $J$=2--1 and $J$=1--0 is almost exactly 1.
When these lines are optically thin and for the
high excitation temperatures deduced in the previous subsection, such
an intensity ratio should approach 4 (the ratio of the squares of the
upper level $J$-value for each transition), which is the opacity ratio
in the high-excitation limit. The measured line ratio of 1 is 
clearly incompatible with optically thin emission. We conclude that 
both $^{12}$CO $J$=1--0 and $J$=2--1 lines are optically thick.  

In our observations the $J$=2--1 line of the isotope $^{13}$CO is significantly weaker
than the line of the main isotope. The emission of this line concentrates
in a few channels around the systemic velocity, between V$_{\rm LSR}$ = 44 to 46 \kms.
By comparing the channel maps of the two lines we conclude that the 
$^{12}$CO/$^{13}$CO $J$=2--1 intensity ratio significantly varies for
the different parts of the nebula. This strong variation can be readily seen from the
total spectra in main-beam brightness units. Values as low as 2.5 are
reached around the peak of emissions at the systemic velocity of the nebula. In
the wings where the emission of the isotope line is much weaker, the
ratio increases rapidly. Such a trend of the line ratio, depending on the intensity, is
expected if the emission of the main isotope is optically thick.  The low
values of the $^{12}$CO/$^{13}$CO intensity ratio near the systemic velocity 
are, on the other hand, too low to represent the abundance ratio, as would be the case if
both lines are optically thin. For instance, the
$^{12}$CO/$^{13}$CO line intensity ratios for circumstellar envelopes around
AGB and post-AGB stars are usually much larger (Bujarrabal et
al.\ 2001), about 10 or higher. We note that similar behavior of 
the $^{12}$CO/$^{13}$CO $J$=2--1 intensity ratio  has been seen in 
a few other cases such as in the molecular envelope around young planetary 
nebula NGC 6302 (Dinh-V-Trung et al. 2008).
\subsection{Estimate of gas temperature}
The kinetic temperature ($T_{\rm k}$) of the molecular gas detected in
our maps can be estimated from the brightness temperature distribution
($T_{\rm mb}$), particularly for the $J$=2--1 transition. We can see
that values as high as $T_{\rm mb}$ $\sim$ 8 K are found at the peak of
the emission in the velocity channel maps around the systemic velocity. 
$T_{\rm mb}$ is approximately equal to $T_{\rm k}$ --
$T_{\rm bg}$ (the cosmic background temperature, 3 K), in the limit of
thermalized level populations, high opacities, and resolved spatial
distribution.  Otherwise, $T_{\rm k}$ must be larger than $T_{\rm mb}$
+ $T_{\rm bg}$, except for very peculiar excitation states. The envelope
is also clearly resolved in our observations.
As we argue in the previous section,
$^{12}$CO $J$=2--1 line is likely optically thick and the gas densities 
are high enough to thermalize the CO low-J lines.
Therefore, we conclude that the kinetic temperature in the
molecular gas in IRAS 08544$-$4431 is typically $\sim$ 10 K.

\subsection{Mass of molecular gas in the nebula}

We use the formula of Nyman \& Olofsson (1993) and Chiu et al. (2006) to 
estimate the mass of molecular gas in IRAS 08544$-$4431.  
We have used the $^{13}$CO $J$=2--1 line, assuming optically
thin emission, a typical excitation temperature $T_{\rm ex}$ of 10 K, and a $^{13}$CO relative
abundance of $f_{\rm ^{13}CO}$ = 2 10$^{-5}$ with respect to H$_2$ (Bujarrabal et al. 2001). 
Because the distance to IRAS 08544$-$4431 is still very uncertain, we adopt a distance $D$ = 1 kpc.
\begin{equation}
M = \frac{16\pi k m_{\rm H}}{hcg_{\rm u}A_{\rm ul}f_{\rm ^{13}CO}}\frac{D^2}{A_{\rm e}}
I_{\rm ^{13}CO}\,Q(T_{\rm ex})\,e^{E_{\rm u}/kT_{\rm ex}}
\end{equation}
Where $A_{\rm e}$ is the effective temperature of the telescope, $Q(T_{\rm ex})$ is
the partition function. The integrated intensity $I_{\rm ^{13}CO}$ of $^{13}$CO J=2--1 measured from
our data and converted to the same angule resolution of the SEST telescope 
is $\sim$0.86 K\kms. That corresponds to a 
molecular gas mass of 4.7 10$^{-3}$ M$_\odot$. As shown in the above equation, the estimated
gas mass scales with square of the distance to the star, which is currently not known. Thus
the gas mass in the evelope will be lower if the star is located at a smaller distance. If we adopt higher excitation
temperature of 250 K as in Maas et al. (2003), the molecular gas mass is $\sim$0.03 M$_\odot$.
The molecular gas mass estimate from Maas et al. (2003) is $\sim$0.02 M$_\odot$ (the larger value
quoted in their paper is probably a typo error), which is a factor
of four higher than our estimate here, mainly due to the difference in the assumed excitation
temperature of the gas.
\section{Implications of our observations}
As seen in section 3, the nebula around IRAS 08544$-$4431 consists of a central component with
strong CO emission but with very low expansion velocity traced mainly by the
$^{13}$CO J=2--1 emission, and a weak bipolar outflow oriented roughly
in the East - West direction. The very narrow
linewidth of the emission from the central component implies an expansion velocity
of only $\sim$1 to 3 \kms. Such low expansion velocity is rarely seen in normal AGB or post-AGB stars where
the expansion velocity is usually in the range of 10 to 20 \kms.
An alternative interpretation commonly put forward in the literature (Maas et al. 2003) is 
that the gas is not in expansion but is distributed
in the form of a stable rotation disk around the central star. At large radial
distance from the star, say a few thousands AU, the rotation velocity is expected to be
quite small, approximately 1 \kms or less for a 1M$_\odot$ central post-AGB star. 
In addition, the spatially extended emitting 
region suggests that such a disk
is seen at moderate inclination angle, i.e. close to face-on. Thus the apparent narrow linewidth 
is consistent with the hypothesis of a rotating disk around the central
post-AGB star of IRAS 08544$-$4431. 
The blueshifted and redshifted wings of $^{12}$CO emission in IRAS 08544$-$4431 most probably comes from
a pair of wide opening angle outflow. Assuming that the bipolar outflow is oriented perpendicular
to the disk, then the whole nebula must be viewed close to the symmetry axis.

The complex kinematics of the molecular gas in IRAS 08544$-$4431 strongly resembles that observed in 
another post-AGB star 89 Her, which is also a prominent member of the class of post-AGB stars with near
infrared excess. The $^{12}$CO emission from 89 Her also 
consists of a central compact core and a pair of
wide opening angle outflow (Bujarrabal et al. 2007). From the compact size of the central core
seen in both radio and near IR interferometric observations, Bujarrabal et al. (2007) suggest that the central 
core is likely to be a rotating disk seen close to face-on, consistent with its very narrow $^{12}$CO emission.
 
One common feature among these peculiar post-AGB stars is that they all have a binary companion. The orbital
period of these systems is relatively short, approximately 1 yr (Maas et al. 2003, de Ruyter et al. 2006). Such close-in 
binary companion may have important dynamical effect because the companion might capture or redirect the
wind material toward the orbital plane, producing a disk-like structure. The presence of a binary companion
and the very pecular kinematics, i.e. the narrow linewidth of CO emission, are
consistent with the interpretation that the gas resides in a stable rotating disk around the post-AGB star. 
In addition, the near infrared excess is particularly prominent
in Red Rectangle, 89 Her and IRAS 08544$-$4431 (De Ruyter et al. 2006). By fitting the SED of these stars de Ruyter et al.
(2006) infer that the dust must be quite hot and must exist within a few tens of stellar radii.
Because there is no evidence for present day dust formation around these hot post-AGB stars, the dust particles must
be stored in a long lived reservoir around the stars, i.e. a rotating disk.
However, to date only the disk around Red Rectangle has been imaged and clearly resolved using radio 
interferometry (Bujarrabal et al. 2005). In this case,
the kinematics of the disk is found to be quite complex with the inner part in Keplerian rotation and some slow
expansion in the outer part. 

Currently, the formation mechanism of the rotating disk around these post-AGB stars is not well understood. 
The source that provides the large amount of angular momentum of the gas and dust
in the rotating disk has not been identified. It has commonly been thought that the gas and dust acquire the
necessary angular momemtum through gravitational interaction with a binary companion, although how that process 
happens has not been demonstrated clearly in detail.
Recently, Akashi \& Soker (2007) propose a different scenario for the formation of these circumbinary rotating disk. They argue that
the interaction between a wide angle jet and the slowly expanding envelope pushes back the
wind material toward the center of the nebula. The back-flowing material will then form a rotating disk
with diameter up to 10$^3$ AU around the central post-AGB star. This model, however, is still rather speculative
and has not been tested in any real three-dimensional hydrodynamic simulation. 

In the near future high angular resolution obserations with ALMA will allow us to resolve the 
central component and to study its spatial kinematics in much greater detail.
\acknowledgments

We are grateful to an anonymous referee for constructive comments which help improve the presentation
of the paper. We thank the SMA staff for their help with the observations. This research has made use of 
NASA's Astrophysics Data System Bibliographic Services
and the SIMBAD database, operated at CDS, Strasbourg, France.

\begin{figure*}[ht]
\plotone{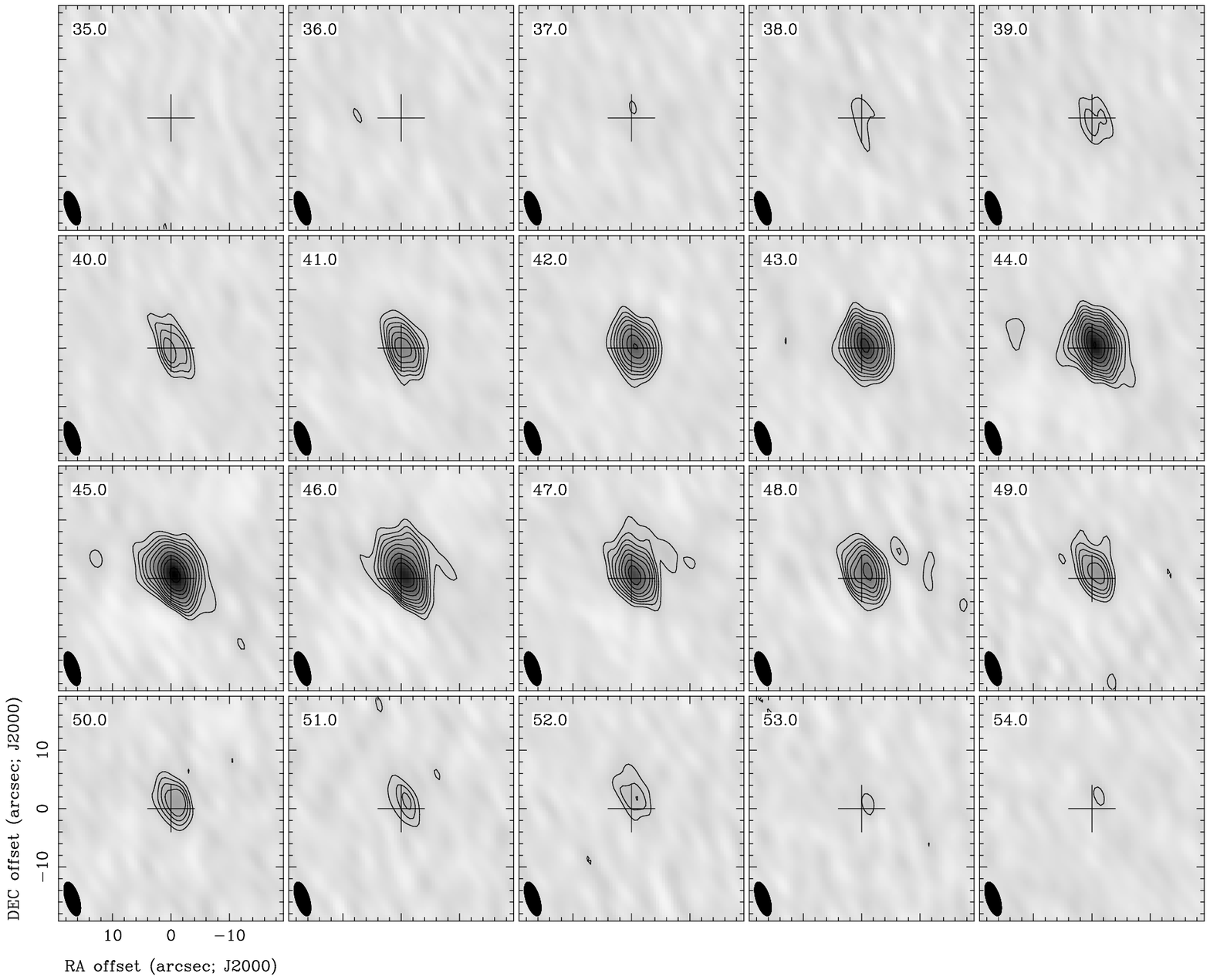}
\caption{Channel maps of $^{12}$CO J=2--1 emission from IRAS 08544$-$4431 shown in both contours and greyscale.
The synthesized beam of 5".6x2".8 is shown in the lower left of each frame. The LSR velocity of each channel
is shown in the upper left. The cross denotes the pointing center of our observations. 
Contour levels are (3, 5, 7, 9, 12, 15, 20, 25, 30, 35, 40)$\sigma$, where $\sigma$ = 0.11
Jy beam$^{-1}$ is the rms noise level.}
\label{fig1}
\end{figure*}

\begin{figure*}[ht]
\plotone{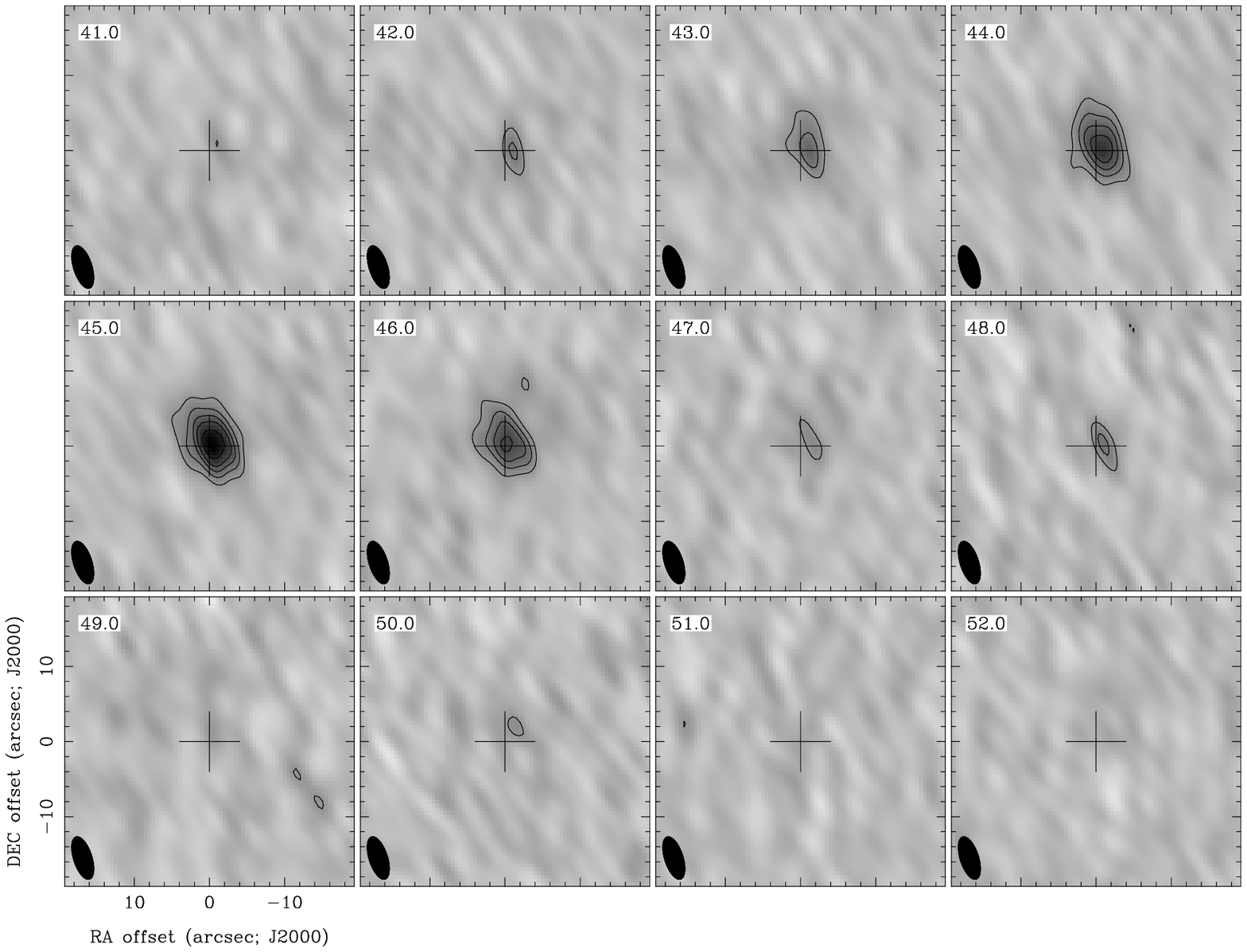}
\caption{Channel maps of $^{13}$CO J=2--1 emission from IRAS 08544$-$4431 shown in both contours and greyscale.
The synthesized beam of 5".6x2".8 is shown in the lower left of each frame. The LSR velocity of each channel
is shown in the upper left. The cross denotes the pointing center of our observations.
Contour levels are (3, 5, 7, 9, 12, 15, 20, 25, 30, 35, 40)$\sigma$, where $\sigma$ = 0.11
Jy beam$^{-1}$ is the rms noise level.}
\label{fig2}
\end{figure*}

\begin{figure*}[ht]
\plotone{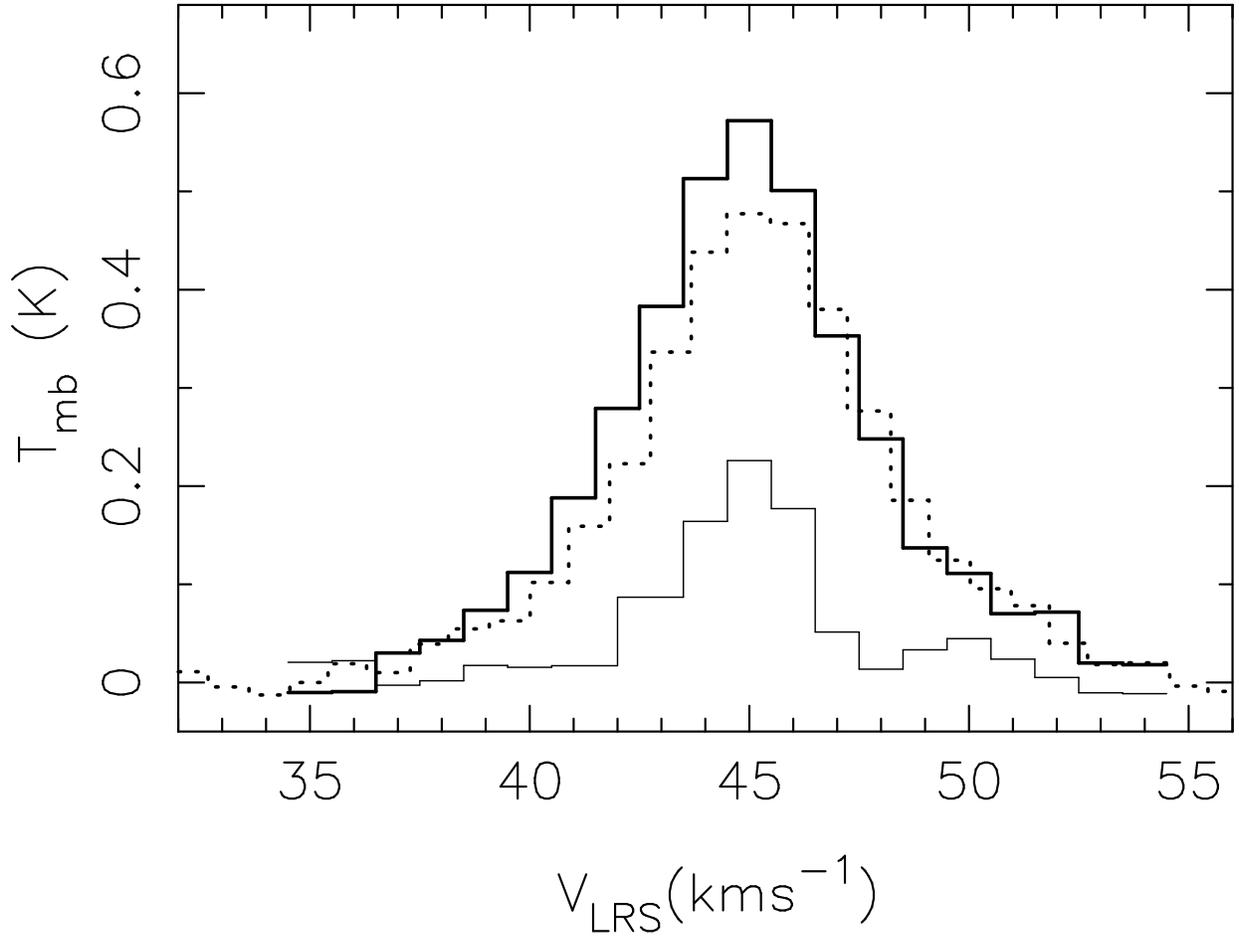}
\caption{CO line profiles. The thick solid line represents the $^{12}$CO J=2--1 emission obtained with the SMA.
The dotted line shows the same line observed with the SEST telescope (Maas et al. 2003). The thin line represents
the $^{13}$CO J=2--1 emission obtained with the SMA. The SMA data have been convolved to the same resolution of 23"
of the SEST telescope.}
\label{fig3}
\end{figure*}

\begin{figure*}[ht]
\setlength{\unitlength}{1cm}
\begin{picture}(16.,18.)
\put(0.0,0.0){\includegraphics*{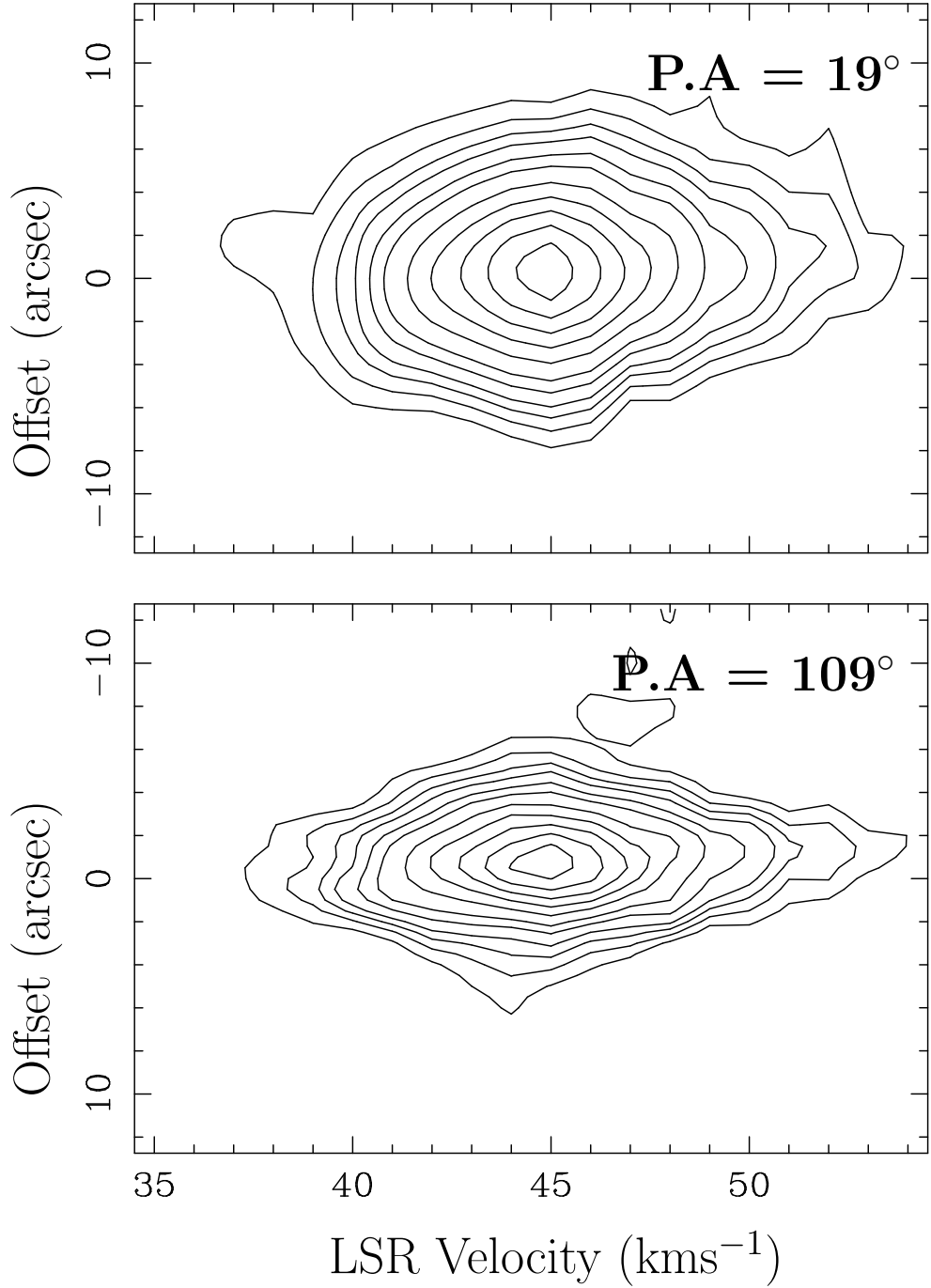}}
\end{picture}
\caption{Position velocity diagrams of $^{12}$CO J=2--1 emission along position angles of 19$^\circ$ (upper frame) and
109$^\circ$, respectively. Contour levels are (3, 5, 7, 9, 12, 15, 20, 25, 30, 35, 40)$\sigma$, where $\sigma$ = 0.11
Jy beam$^{-1}$ is the rms noise level.}
\label{fig4}
\end{figure*}

\end{document}